\documentclass[prb,preprintnumbers,twocolumn,amsmath,amssymb,superscriptaddress]{revtex4}
\usepackage{graphicx}
\usepackage{dcolumn}
\usepackage{bm}
\usepackage{psfrag}
\newcommand{\qvec}{{\bf q}}

\usepackage{color}

\begin{document}
\title{Phase separation and long wave-length charge instabilities in spin-orbit coupled systems} 
\author{G. Seibold}  
\affiliation{Institut f\"ur Physik, BTU Cottbus-Senftenberg, 
PBox 101344, 03013 Cottbus, Germany}
\author{D. Bucheli}
\affiliation{Dipartimento di Fisica 
Universit\`a di Roma Sapienza, Piazzale Aldo Moro 5, I-00185 Roma, Italy}
\author{S. Caprara}
\author{M. Grilli}
\affiliation{Dipartimento di Fisica 
Universit\`a di Roma Sapienza, Piazzale Aldo Moro 5, I-00185 Roma, Italy}
\affiliation{
ISC-CNR and Consorzio Nazionale Interuniversitario per le Scienze 
Fisiche della Materia, Unit\`a di Roma Sapienza, Italy}
\date{\today}

\begin{abstract}
We investigate a two-dimensional electron model with Rashba spin-orbit 
interaction where the coupling constant $g=g(n)$
depends on the electronic density.  It is shown that this dependence may
drive the system unstable towards a long-wave length charge density wave (CDW) where the
associated second order instability occurs in close vicinity to global
phase separation. For very low electron densities the CDW instability is 
nesting-induced and
the modulation follows the Fermi momentum $k_F$. At higher density 
the instability criterion
becomes independent of $k_F$ and the system may become unstable in a broad
momentum range. Finally, upon filling the upper spin-orbit split band,
finite momentum instabilities disappear in favor of phase separation alone.
We discuss our results with regard to the inhomogeneous 
phases observed at the LaAlO$_3$/SrTiO$_3$ or LaTiO$_3$/SrTiO$_3$ interfaces.
\end{abstract}

\maketitle
\section{Introduction}
The observation of a high-mobility electron gas at the
interface of LaAlO$_3$/SrTiO$_3$ \cite{ohtomo} has attracted great interest  
due to the variety of possibilities of controlling and modifying the
properties of a two-dimensional (2D) electron 
system \cite{mannhart08,heber09,mannhart10,hwang12}.
Due to the large dielectric constant of the SrTiO$_3$ substrate,
the carrier density and mobility of the 2D electron gas can be tuned by
applying a gate voltage either across the substrate \cite{caviglia08,bell09} or,
in a less easy implementation, by depositing the gate on top of the 
thin upper LaAlO$_3$ layer.
Such electric field effect can be used in order to modulate the electron density and hence
the superconductivity \cite{reyren,espci1}. More recently it has been shown by magnetotransport
experiments \cite{biscaras12} that the
electron gas in LaTiO$_3$/SrTiO$_3$ heterostructures is composed
of two types of carriers, a majority of low mobility carriers that
are always present and high-mobility ones which induce superconductivity
upon injection with positive gating.
In the latter regime, an analysis of the resistivity $R$ has revealed
pronounced tails in the $R$ {\it vs.} temperature curves \cite{caviglia08,caprara11}. These tails can be understood with the assumption of an
inhomogeneous superconducting order parameter distribution \cite{caprara11,bucheli13,caprara13}. Naturally, the low dimensionality of the electron gas
emphasizes the role of disorder which unavoidably is created during the growing process, like, e.g.,
oxygen vacancies. However, it has been proposed \cite{caprara12,bucheli132} that the strong Rashba spin-orbit 
(RSO) interaction may also provide an
intrinsic mechanism for the occurrence of electronic phase separation (PS).
This point of view is supported by the observation of a negative 
compressibility \cite{li11} in strongly charge depleted LaAlO$_3$/SrTiO$_3$ 
interfaces where the effect seems to go beyond the standard contribution
from exchange interactions in a 2D electron gas.

PS together with frustration due to disorder would then be responsible
for the formation of mesoscopic superconducting islands embedded in a (weakly localizing) metallic
background. The mutual phase coherence between neighboring islands would
then generate a low-dimensional percolative superconducting network
which is compatible with the observation of 'tailish' $R(T)$ curves \cite{bucheli13,caprara13}.

From a formal point of view, the investigation of Ref. \cite{caprara12} 
was restricted to an analysis of the chemical potential {\it vs.} electron
density, and it was subsequently extended
investigating the role of temperature and magnetic field
in restoring the homogeneous electronic state \cite{bucheli132}. 
It has been shown that upon increasing the RSO a negative slope
corresponding to a PS instability can be induced which is equivalent to
a divergence of the density-density response function at zero transferred
momentum. Naturally the question arises whether this mechanism is more general
and can even lead to finite momentum instabilities. It is exactly
this problem which we intend to analyze in the present paper.

In this context it has been shown recently \cite{scheurer} 
within a multiband model, 
that scattering processes between the nested regions of Rashba split subbands 
may induce a variety of competing phases. In case of a conventional 
(attractive) pairing these include also charge density waves competing
with conventional superconductivity.
Here we will demonstrate that charge-density wave instabilities can also
be driven by a density-dependent RSO interaction which for 
the sake of simplicity will be analyzed within a one-band model.

\section{Model and Formalism}\label{sec:model}
In order to describe the electronic structure of the interface 
electron gas we consider a 2D lattice model 
with strong Rashba coupling described by the Hamiltonian \cite{evangelou87,ando89}

\begin{eqnarray}                                         
H&=&\sum_{ij\sigma}t_{ij}c^\dagger_{i\sigma}c_{j\sigma} 
+\sum_i \left\lbrack \gamma_{i,i+y} j_{i,i+y}^x - \gamma_{i,i+x} j_{i,i+x}^y
\right\rbrack\nonumber \\ &+& \sum_{i,\sigma} \lambda_i \left\lbrack
c^\dagger_{i\sigma}c_{i\sigma} - n_i \right\rbrack  \label{eq:ham}
\end{eqnarray}   
with $n_i \equiv \sum_\sigma \langle c_{i\sigma}^\dagger c_{i\sigma}\rangle$. 
Here, the first term describes the kinetic energy of electrons on a 
square lattice (with lattice constant $a$)
and we only take hopping between nearest-neighbors into account ($t_{ij}
\equiv -t$ for $|R_i-R_j|=a$, and $t_{ij}=0$ otherwise). The second term is the RSO coupling, 
where
\begin{equation}\label{eq:spcurr}
 j_{i,i+\eta}^\alpha = -i\sum_{\sigma\sigma'}\left\lbrack c^\dagger_{i\sigma} \tau^\alpha_{\sigma\sigma'} 
 c_{i+\eta,\sigma'} - c^\dagger_{i+\eta,\sigma} \tau^\alpha_{\sigma'\sigma} c_{i,\sigma}
\right\rbrack  
\end{equation}
denotes the $\alpha$-component ($\alpha=x,y,z$) of the spin-current flowing on the bond
between $R_i$ and $R_{i+\eta}$, and $\tau^\alpha$ 
are the Pauli matrices. The coupling constants obey the
relation $\gamma_{ij}=\gamma_{ji}$. The last term in Eq. \ref{eq:ham} ensures the local density
fixed to $n_i$ via the local chemical potential $-\lambda_i$ which will
become important below when we deal with density-dependent RSO couplings. 

Note that for homogeneous coupling constants $\gamma_{i,i+\eta}\equiv \gamma$ 
the  RSO term takes the usual form in momentum space

\begin{equation}
H^c=\sum_{\bm k\sigma\sigma'}\left\lbrack \alpha_{\bm k}^x \tau^x_{\sigma\sigma'}
+ \alpha_{\bm k}^y \tau^y_{\sigma\sigma'}\right\rbrack c_{\bm k\sigma}^\dagger c_{\bm k\sigma}
\end{equation}
where
${\bf \alpha}_{\bm k} = \gamma (
-\frac{\partial \epsilon_{\bm k}}{\partial k_y},
\frac{\partial \epsilon_{\bm k}}{\partial k_x},0)$
and $\epsilon_{\bm k}$ is the 2D lattice dispersion law. Thus a
constant $\gamma$  reproduces the customary 
$(\vec{\bm v}\times \vec{\bf \sigma})$ form of the RSO interaction.

The RSO coupling constants depend on the electric field $\bm E$ perpendicular to the
2D electron gas. Following Refs. \cite{caprara12} and \cite{bucheli132}, we argue that the electric field
is determined by the charge density in the layer thus yielding a dependence
of the RSO coupling constant on the charge density.
For the present purpose, it is important to note that this dependency
is a local one, i.e., a local charge fluctuation will locally alter
the electric field and thus also locally change the RSO coupling
constant.
We therefore set $\gamma_{i,i+\eta}=\gamma_{i,i+\eta}(n_i, n_{i+\eta})$
and since by symmetry $\partial\gamma_{i,i+\eta}/\partial n_i=
\partial\gamma_{i,i+\eta}/\partial n_{i+\eta}$ we  
consider the following coupling
\begin{equation}\label{eq:coup}
\gamma_{i,i+\eta}= a_0 + a_1 (n_i + n_{i+\eta}).
\end{equation}
Note that in Ref. \cite{caprara12} a more general form of the
density-dependent coupling has been considered which saturates
at large densities. Here we abstain from this extension, in order
to keep the number of parameters to a minimum. However, the following
instability analysis can be straightforwardly extended to general
density-dependent RSO couplings which will induce the
formation of electronic inhomogeneities and thus concomitant
variations in the local chemical potential.
The latter can be obtained self-consistently by minimizing the
energy with respect to the local density $\partial E/\partial n_i =0$
which yields
\begin{equation}\label{eq:sc}
\lambda_i = a_1 \left(\langle j_{i,i+y}^x\rangle
- \langle j_{i,i+x}^y\rangle\right) .
\end{equation}
A few words are now in order to account for the introduction of the Lagrange multiplier field 
$\lambda_i$. In general the thermodynamic properties of fermionic systems are more easily 
calculated in the grand-canonical ensemble, defined by the potential 
$\Omega=F-\mu N$, where $F$ is the free energy and $N$ the total particle number of the system. 
In this ensemble the chemical potential is fixed while the particle density (per unit cell)
$n=N/L^2$ is a fluctuating 
quantity ($L^2$ is the number of lattice sites). Therefore, when we calculate the chemical potential 
via the density-dependent grand-canonical potential, the density [and therefore also $\gamma(n)$]
is not known a priori. An elegant way to circumvent this problem is the introduction of a Lagrange 
multiplier $\lambda$ as was done in Ref. \cite{bucheli132}. In the present non-homogeneous case 
we proceed similarly and we take the energy to be dependent on a real parameter $n_i$, determine the 
local chemical potential $\mu(\lambda_i,n_i)$, and 
then chose $\lambda_i$ to impose $n_i=\langle  c^\dagger_{i\sigma}c_{i\sigma} \rangle$ via Eq. (\ref{eq:sc}).

If we insert the Lagrange parameter Eq. (\ref{eq:sc}) in the Hamiltonian 
Eq. \ref{eq:ham} the energy functional reads as
\begin{eqnarray}
E &=& E_0
+a_1 \sum_{i}\left(n_i + n_{i+y}\right)
\langle j_{i,i+y}^x\rangle \nonumber \\
&-&a_1 \sum_{i}\left(n_i + n_{i,i+x}\right)
\langle j_{i,i+x}^y\rangle  \label{eq:exp}
\end{eqnarray}
with 
$$
E_0=\sum_{ij\sigma}t_{ij}\langle c^\dagger_{i\sigma}c_{j\sigma}\rangle
+ \sum_i \left\lbrack \gamma^0_{i,i+y} \langle j_{i,i+y}^x\rangle - \gamma^0_{i,i+x} 
\langle j_{i,i+x}^y\rangle
\right\rbrack
$$
and the notation $\gamma^0$ refers to the coupling at a given density.

From Eq. (\ref{eq:exp}) it becomes apparent that the density-dependent
coupling induces an effective density-current interaction. 
Assume that the problem has been solved for a given homogeneous density
(in the following, we take densities and currents as site independent).
Then, one can obtain the instabilities of the system from the expansion
of the energy in the small fluctuations of the density matrix in momentum space
\begin{eqnarray}
\delta E&=& Tr(H\delta\rho)  
+\frac{a_1}{L^2}\sum_{\bm q}
\cos(\frac{q_y}{2}) \left\lbrack \delta\rho_{\bm q}\delta j^x_{-\bm q}
+ \delta j^x_{\bm q} \delta\rho_{-{\bm q}}\right\rbrack \nonumber \\
&-&\frac{a_1}{L^2}\sum_{\bm q}
\cos(\frac{q_x}{2}) \left\lbrack \delta\rho_{\bm q}\delta j^y_{-{\bm q}}
+ \delta j^y_{\bm q}\delta\rho_{-{\bm q}}\right\rbrack \label{eq:deltae} .
\end{eqnarray}

The fluctuations are given by
\begin{eqnarray*}
\delta j^x_{{\bm q}} &=& -2t\sum_{\bm k\sigma\sigma'} \sin(k_y+\frac{q_y}{2})
c^\dagger_{\bm k+\bm q,\sigma}\tau^x_{\sigma\sigma'}c_{\bm k,\sigma'}\\
\delta j^y_{\bm q} &=& -2t\sum_{\bm k\sigma\sigma'} \sin(k_x+\frac{q_x}{2})
c^\dagger_{\bm k+\bm q,\sigma}
\tau^y_{\sigma\sigma'}c_{\bm k,\sigma'}\\
\delta \rho_{\bm q} &=& \sum_{\bm k\sigma\sigma'}c^\dagger_{\bm k+\bm q,\sigma}
{\bf 1}_{\sigma\sigma'}c_{\bm k,\sigma'} \,.
\end{eqnarray*}
and the instabilities can now be determined from a standard random phase
approximation (RPA). For this purpose we introduce response functions
$$
\chi_{AB}(\bm q)=-\frac{i}{L^2} \int dt \langle {\cal T}
\delta A_{\bm q}(t) \delta B_{-\bm q}(0)\rangle,
$$
where $\delta A_{\bm q}$ and $\delta B_{\bm q}$ refer to the fluctuations defined above.

The non-interacting susceptibilities can be obtained from the
eigenstates of the Rashba Hamiltonian Eq. (\ref{eq:ham}). 
Denoting the response functions in matrix form
\begin{equation}
\underline{\underline{\chi^{0}({\bm q})}}=
\left(\begin{array}{ccc}
\chi^{0}_{jx,jx} & \chi^{0}_{jx,jy} & \chi^{0}_{jx,\rho} \\
\chi^{0}_{jy,jx} & \chi^{0}_{jy,jy} & \chi^{0}_{jy,\rho} \\
\chi^{0}_{\rho,jx} & \chi^{0}_{\rho,jy} & \chi^{0}_{\rho,\rho}
\end{array}\right)
\end{equation}
and the interaction, derived from Eq. (\ref{eq:deltae}) as
\begin{equation}
\underline{\underline{V({\bm q})}}=
\left(\begin{array}{ccc}
0 & 0 & 2\gamma'\cos(\frac{q_y}{2}) \\
0 & 0 & -2\gamma'\cos(\frac{q_x}{2}) \\
2\gamma'\cos(\frac{q_y}{2}) & -2\gamma'\cos(\frac{q_x}{2}) & 0
\end{array}\right)
\end{equation}
the full RPA response is given by
\begin{equation}\label{eq:rpa}
\underline{\underline{\chi({\bm q})}}=\left(\underline{\underline{\bf 1}} 
- \underline{\underline{\chi^{0}({\bm q})}}
\underline{\underline{V({\bm q})}}\right)^{-1}
\underline{\underline{\chi^{0}({\bm q})}}.
\end{equation}
The instabilities can thus be obtained from the zeros of the determinant
$|{\bf 1} - \chi^{0}({\bm q}) V({\bm q})|$.

Here, the element $\chi_{\rho\rho}(\bm q=0)$ is proportional to the compressibility, 
i.e., within our sign convention, proportional to the inverse of 
$-\partial\mu/\partial n$. A (locally) stable system thus corresponds
to $\chi_{\rho\rho}(\bm q=0)<0$ whereas an unstable system is characterized by
$\chi_{\rho\rho}(\bm q=0)>0$. Analogously, a charge density wave (CDW) instability is
signalled by a divergence in $\chi_{\rho\rho}(\bm q=\bm Q_{CDW}\ne 0)$, where $\bm q=\bm Q_{CDW}$ is 
the momentum which becomes unstable upon entering the symmetry-broken 
CDW phase.
We also note that a non-linear dependency of the RSO coupling on
densities would generate an additional density-density type
contribution to the interaction matrix, i.e., $V_{\rho\rho}({\bm q})$. 
 
\section{Long wave-length CDW's}\label{sec:results}
For the sake of definiteness, in our calculations we fix the constant part of the RSO coupling
in Eq.  \ref{eq:coup} to $a_0/t=0.5$. The stability of the system
is then analyzed in the $(n; a_1/t)$ parameter space where $n$ is the
electron density (per unit cell) and $a_1/t$ is the component of the 
RSO interaction which couples to the density [cf. Eq. (\ref{eq:coup})].

\begin{figure}[ttb]
\includegraphics[width=8cm,clip=true]{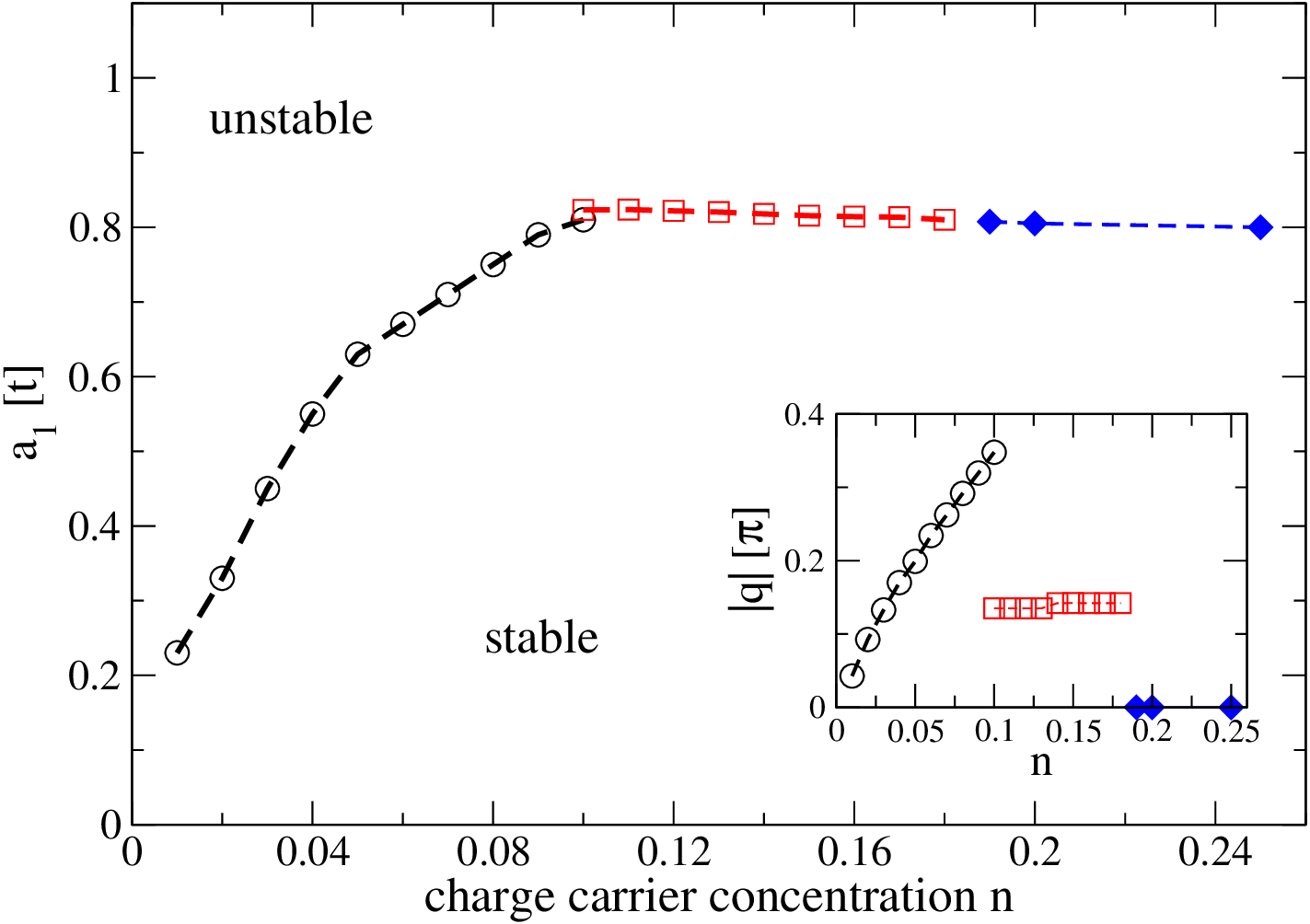}
\caption{Main panel: Instability line in the $a_1/t$ vs. electron density plane with fixed $a_0=0.5t$.
The range where only the lower RSO-split band is occupied corresponds to the
black circles. Above $n=0.105$ both RSO-split bands are occupied (red squares).
Both, circles and squares indicate the transition towards a long-wave length
CDW whereas diamonds refer to a ${\bf q}=(0,0)$ instability.
Inset: Characteristic momentum of the density instability vs. electron density.}
\label{fig1}
\end{figure}

Our results are summarized in Fig. \ref{fig1} which, in the main panel,
reports the instability line. Up to $n\approx 0.1$ (circles) it separates
a homogenous Fermi liquid from a long-wave length CDW where both 
$a_1/t$ and the CDW
modulation $Q_{CDW}$ increase continuously with $n$ (lower right inset).
In the range $0.1 \lesssim n \lesssim 0.18$ (squares) $Q_{CDW}$ drops 
discontinuously 
to a value of $Q_{CDW}\approx 0.14\pi$ and is only weakly density-dependent.
Also the slope of the instability line $a_1/t$ {\it vs.} $n$ 
decreases only weakly with the electron density. 
Finally, one enters a PS regime above $n\approx 0.18$ (diamonds) where the unstable momentum discontinuosly 
jumps to $q=0$ identifying a spinodal line with divergent compressibility. 
This proximity between the finite-$\qvec_c$ instability and the spinodal line indicates that 
long-wavelength CDW fluctuations are naturally accompanied by soft density fluctuations at
$\qvec=(0,0)$  and {\it vice versa}.

\begin{figure}[ttb]
\includegraphics[width=8cm,clip=true]{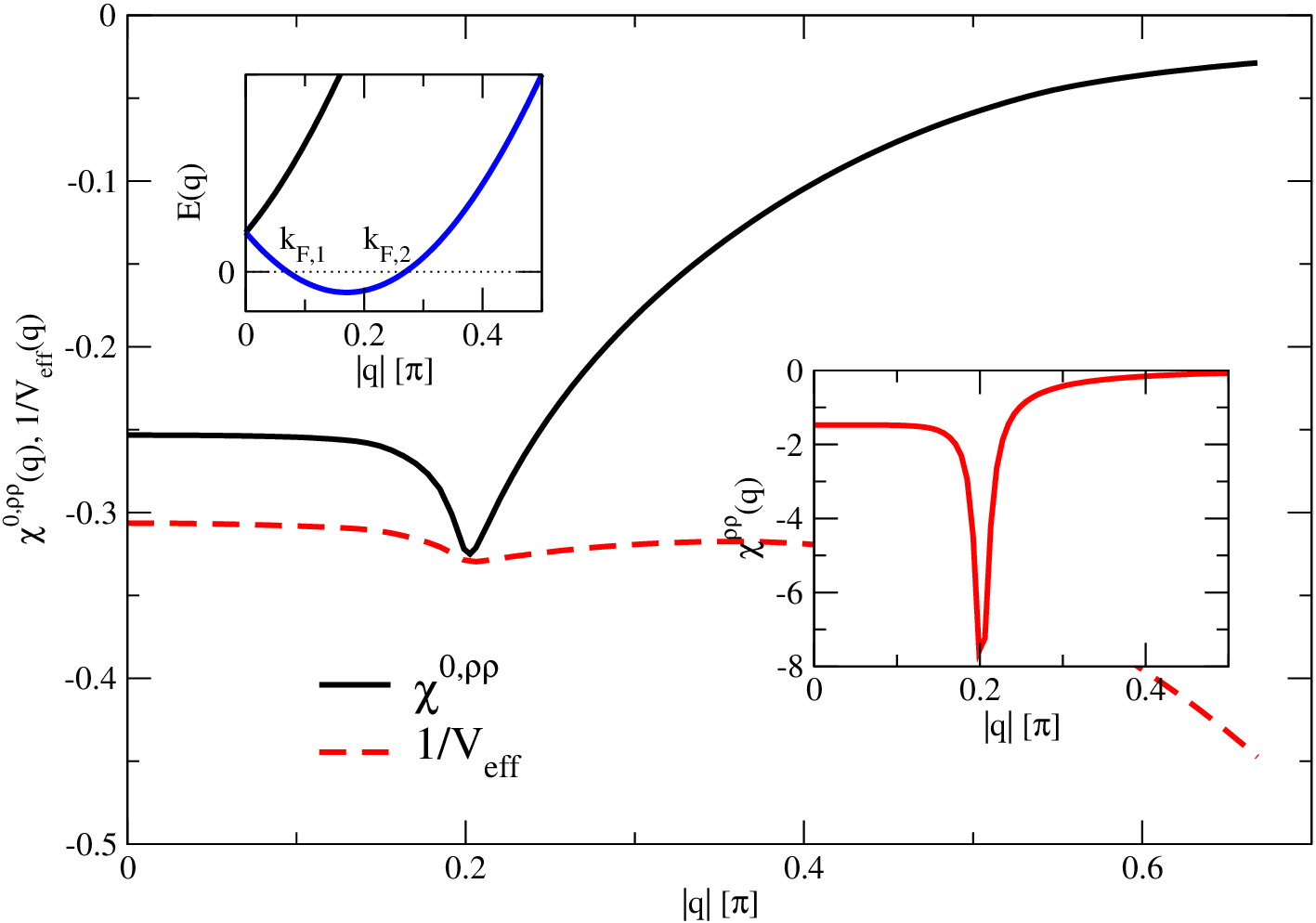}
\caption{Main panel: Non-interacting susceptibility $\chi^{0}_{\rho\rho}({\bm q})$
and inverse effective interaction $1/V_{eff}$ in the low density regime.
The lower right inset shows the full density-density response and 
the upper left inset the bands. Momentum scans are along the $(11)$ direction.
Parameters:  $n=0.05$, $a_0=0.5t$, $a_1=0.62t$.}
\label{fig2}
\end{figure}

In order to analyze the structure of $\chi_{\rho\rho}({\bm q})$ in more detail,
it is instructive to define an effective density-density 
interaction
\begin{equation}
V_{eff}({\bm q})=\frac{1}{\chi^{0}_{\rho\rho}({\bm q})}-\frac{1}{\chi_{\rho\rho}({\bm q})},
\end{equation}
which is a function of the non-interacting susceptibility
$\chi^{0}_{\rho\rho}({\bm q})$
and of the fully dressed $\chi_{\rho\rho}$.  According to Eq. (\ref{eq:rpa}),
this latter includes all the current-current and mixed current-density processes. 
Therefore $V_{eff}({\bm q})$ describes the density-density interaction
effectively mediated by all different fluctuations. The 
correlation function $\chi_{\rho\rho}({\bm q})$ is then formally
determined from a standard RPA-like expression
\begin{equation}
\chi_{\rho\rho}({\bm q})=\frac{\chi^{0}_{\rho\rho}({\bm q})}{1-V_{eff}({\bm q})\chi^{0}_{\rho\rho}({\bm q})}\,.
\end{equation}
Fig. \ref{fig2} reports $\chi^{0}_{\rho\rho}({\bm q})$ together with 
$1/V_{eff}({\bm q})$ in the main panel for $n=0.05$. For small momenta,
the static density-density correlation function is characterized by a flat part
which extends to $2k_{F,1}$ corresponding to scattering between 
opposite states within the lower RSO-split band (cf. upper inset in Fig. 
\ref{fig2}). On the other hand, $\chi^{0}_{\rho\rho}({\bf q})$ is strongly 
enhanced for momenta which connect the two different branches of the 
Fermi surface of the lower RSO-split band (i.e., $q=k_{F,2}-k_{F,1}$)
which drives the instability
in this density range. Since we are considering a lattice model
small anisotropies in the correlation functions appear already
at low density and favor the appearance of the instability along
the diagonal $(11)$ direction. The full density-density correlation
function, which is reported in the lower right inset to Fig. \ref{fig2},
is then characterized by a sharp peak at $Q_{CDW}$ which shifts
to larger momenta with increasing density. 

\begin{figure}[ttb]
\includegraphics[width=8cm,clip=true]{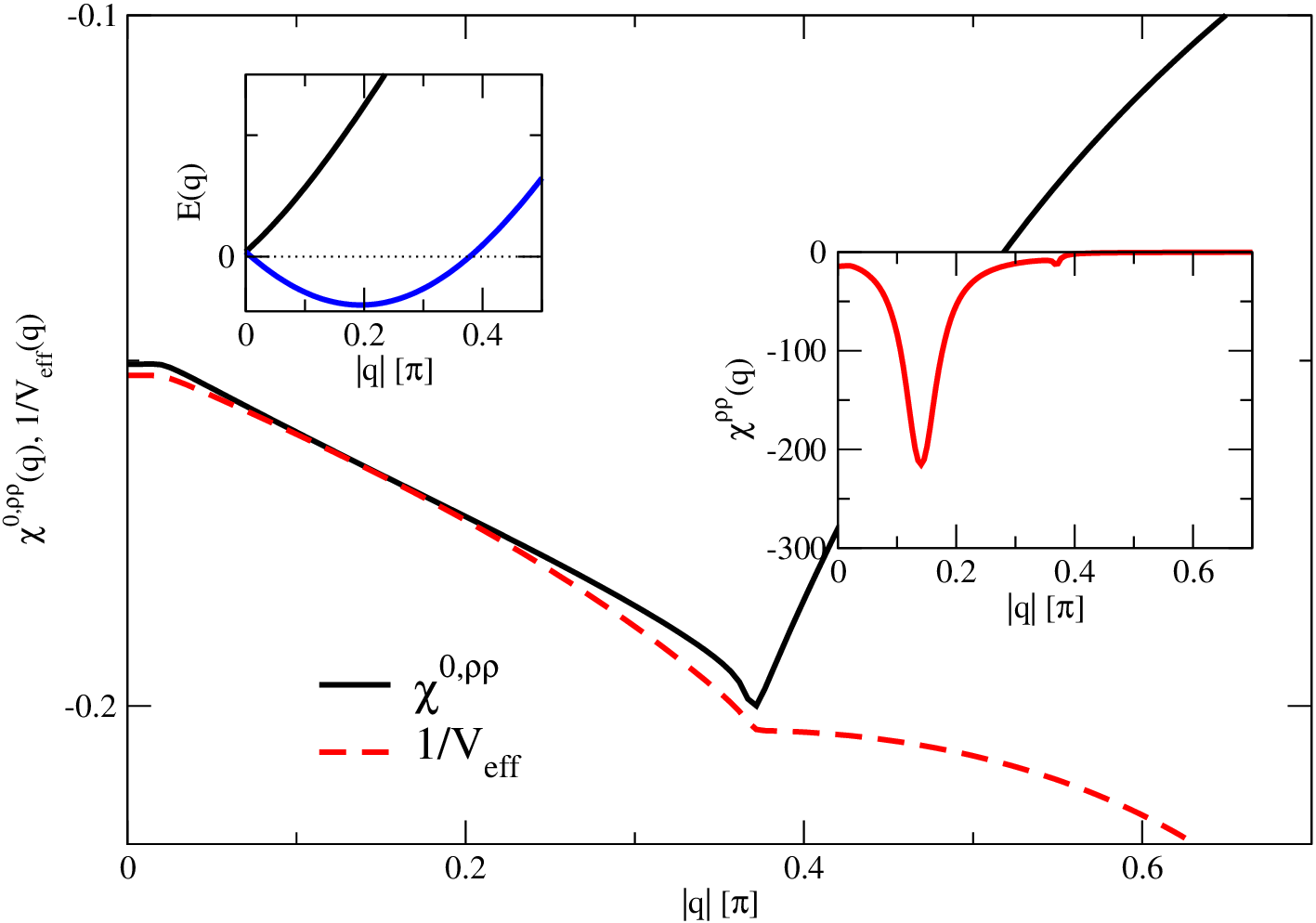}
\caption{Same as Fig. \ref{fig2} but for 
parameters $n=0.12$, $a_0=0.5t$, $a_1=0.822t$.
Momentum scans are along the $(10)$ direction.}
\label{fig4}
\end{figure}

Above $n\approx 0.1$ the nesting-induced CDW  is replaced
by an incommensurate CDW instability at smaller momenta. The situation
is analyzed in Fig. \ref{fig4} for $n=0.12$. Due to the
reduced value of $k_{F,1}$ the constant part of $\chi^{0}_{\rho\rho}({\bf q})$
is now limited to small momenta beyond which enhanced
interband scattering induces an almost linear increase of the charge density
susceptibility. The nesting peak at $q=k_{F,2}-k_{F,1}$ is now much
less pronounced since the curvatures of the Fermi surfaces at  
$k_{F,2}$ and $k_{F,1}$ become more different with increasing density.
As a consequence $1/V_{eff}$ touches $\chi^{0}_{\rho\rho}({\bf q})$
in the quasi-linear regime which causes the appearance of a
broad peak in the full charge correlation function (right inset
to Fig. \ref{fig4}) whereas the nesting peak is only visible
as a small feature at $q\sim 0.38 \pi$. The unstable momentum hardly changes with $n$
and is located along the vertical (horizontal) direction of the Brillouin zone.

\begin{figure}[thb]
\includegraphics[width=8cm,clip=true]{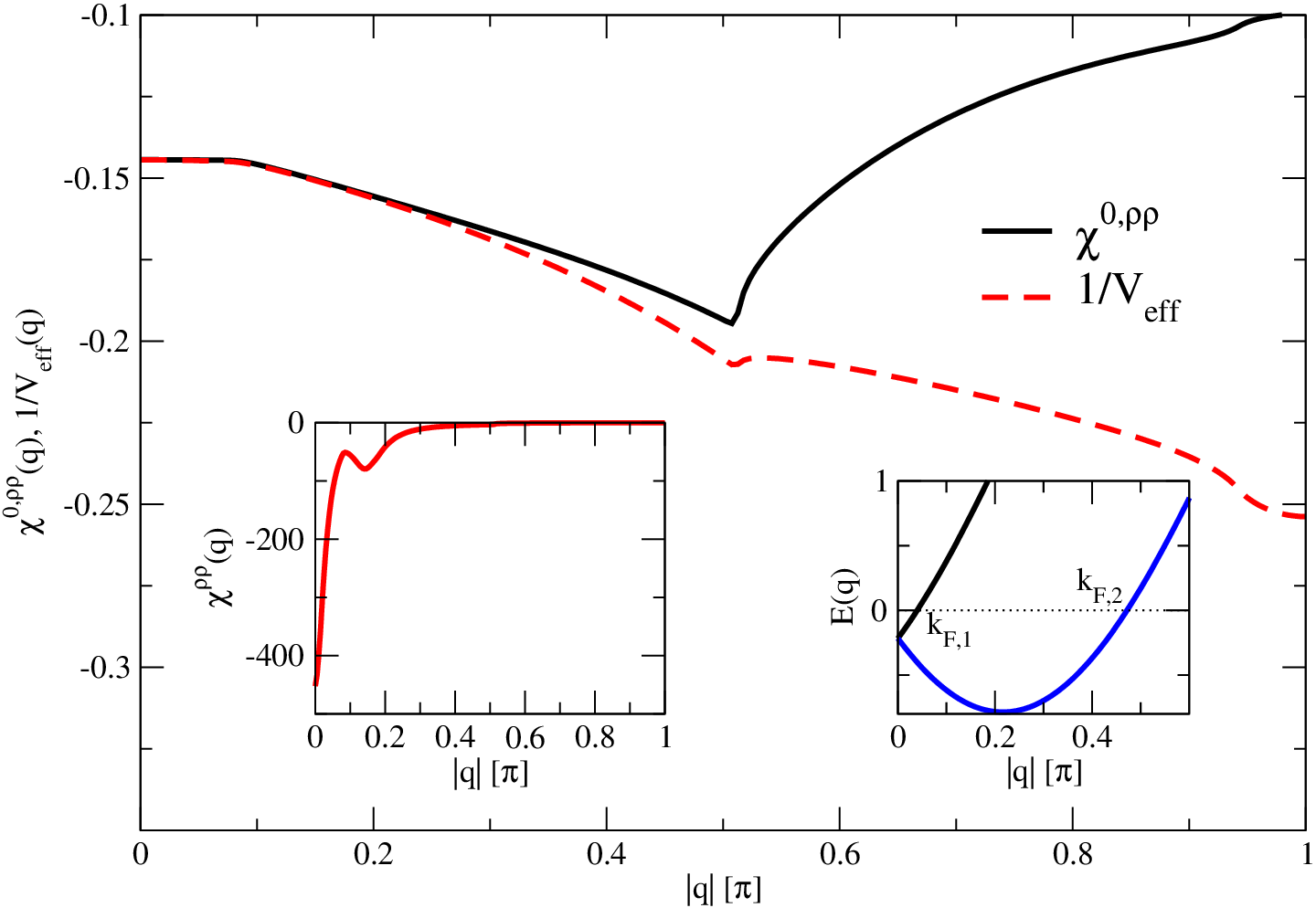}
\caption{Same as Fig. \ref{fig2} but for parameters 
$n=0.19$, $a_0=0.5t$, $a_1=0.8075t$.}
\label{fig5}
\end{figure}

Finally, when the chemical potential enters the upper RSO split 
band for $n \gtrsim 0.18$,  another discontinuous transition occurs
towards a $q=0$ instability. The bare charge correlation function
$\chi^{0}_{\rho\rho}({\bm q})$ (cf. main panel of Fig. \ref{fig5}) has
as similar structure than in the previous case (Fig. \ref{fig4}), however,
the flat part for small momentum extends now towards twice the
Fermi momentum of the upper band $2 k_{F,1}$ (cf. lower right inset 
to Fig. \ref{fig5}). The peak in $\chi^{0}_{\rho\rho}({\bm q})$ is determined
by interband scattering between Fermi momenta of the RSO split
bands and thus occurs at $k_{F,2}-k_{F,1}$. The full charge correlations
$\chi_{\rho\rho}({\bm q})$ (lower left inset to Fig. \ref{fig5}) still
shows an enhancement at the incommensurate momentum $Q_{CDW} \approx 0.2\pi$, 
however, the divergence is now clearly shifted to $q=0$. 
As analyzed further below, this instability corresponds
to a local maximum which develops in $\mu(n)$ upon increasing
$a_1$, so that the static compressibility 
$\kappa=(\partial\mu/\partial n)^{-1}$ 
diverges.

\section{Phase separation}\label{apa}
The occurrence of electronic PS is locally signaled by
the divergence of the compressibility $\kappa=(\partial\mu/\partial n)^{-1}$
marking the spinodal instability line. 

\begin{figure}[thb]
\includegraphics[width=8cm,clip=true]{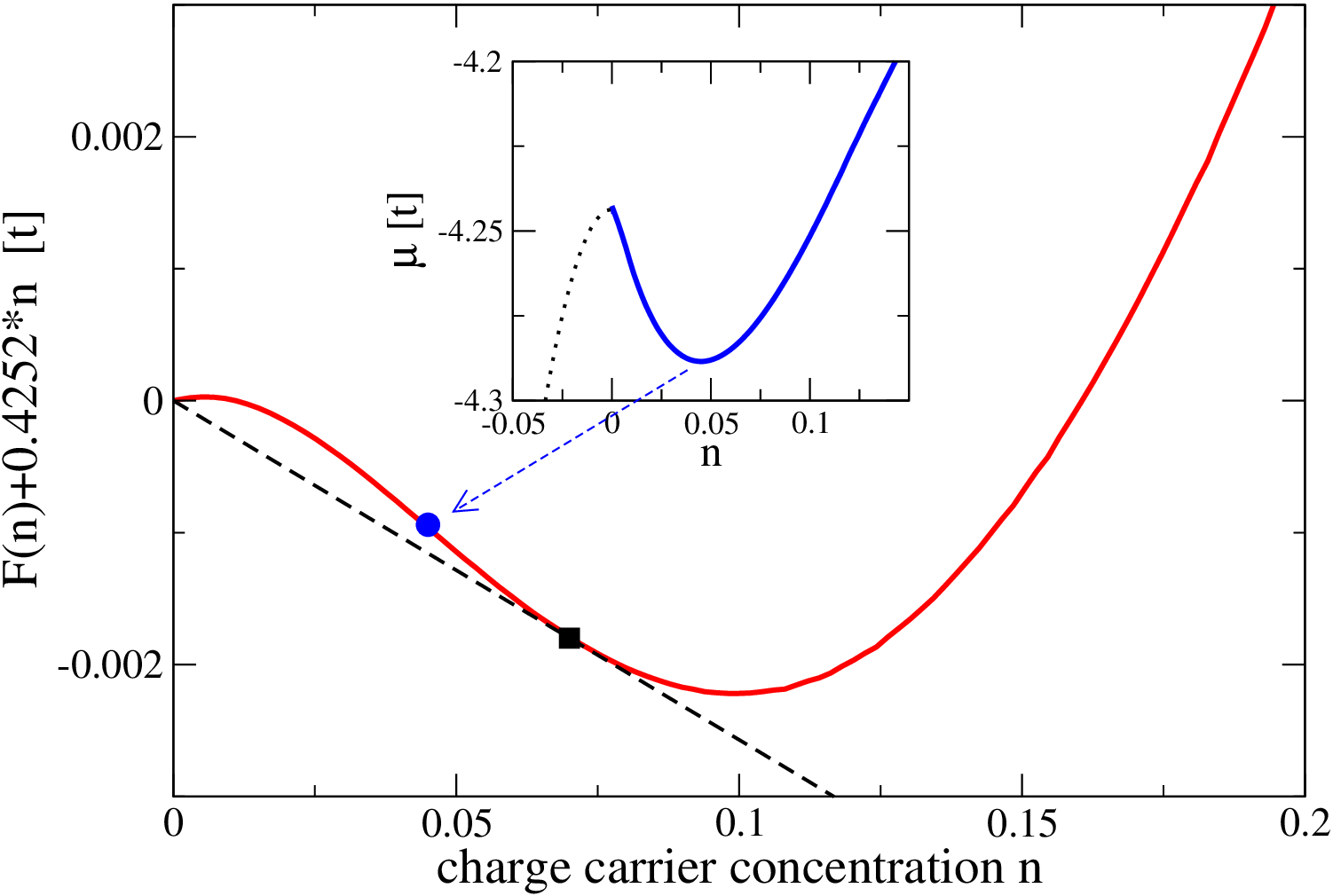}
\caption{Main panel: Free energy $F(n)$ as a function of electron
density (solid). The dashed line reports the Maxwell
construction and the point of contact (square) with $F(n)$ indicates the
global transition to PS. 
Inset: Chemical potential as a function of electron
density. The minimum, which indicates the local instability
point towards PS, corresponds to the point of
inflection in $F(n)$ and is indicated by a circle. The dotted line
sketches the the behavior in case of a multiband structure as discussed
in the text. Parameters: $a_0=0.5t$, $a_1=0.7t$.}
\label{fig6}
\end{figure}

\begin{figure}[thb]
\includegraphics[width=8cm,clip=true]{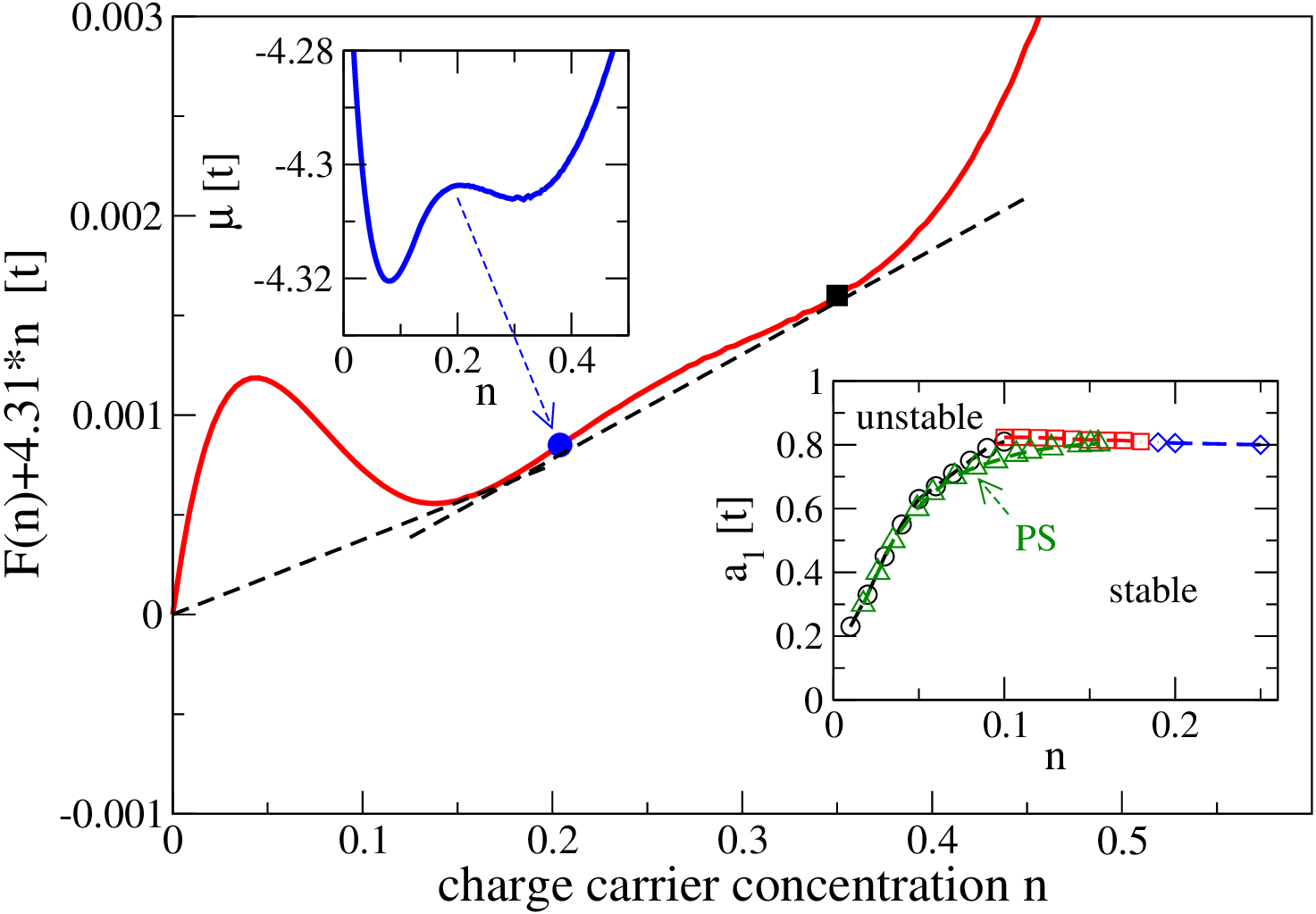}
\caption{Main panel and upper left inset: Same as Fig. \ref{fig6} but for 
larger coupling $a_1=0.804 t$.
The upper right inset shows the PS instability (green triangles) 
together with the 
transitions towards CDW order as already reported in Fig. \ref{fig1}.}
\label{fig7}
\end{figure}

However, the real
PS region is found by implementing the Maxwell construction, which identifies
the broader parameter region where the system no longer is formed
by a single homogeneous phase. The most natural way to 
perform the Maxwell construction is by the tangent construction
to the free energy, which bypasses the unstable region where the free energy has a wrong
concavity, finding a region with lower free energy realized as a linear superposition
of two stable phases with different densities.
However, to implement this procedure when various phases are
present, one should know the free energy of the 
various phases and refer to the lowest one in each region of the phase
diagram.  Unfortunately, this program is rather difficult
when incommensurate phases are present, which make the calculation of the
(meta)stable phases a non-trivial task. Therefore, we will
investigate the impact of the Maxwell construction on the phase diagram, by  
only considering the free energy of the uniform phase. We are aware that 
at low density the stable phases have an incommensurate CDW order (order
parameter $\Delta$), which lowers the free energy by $\sim \Delta^2/a_1$. 
This would of course modify the boundaries of the
Maxwell construction, which should determine the PS between
a uniform phase at high density and a CDW phase at low density.

Nevertheless we find it instructive to perform the Maxwell construction
in this approximate way to extract still valuable estimates on the size and
structure of the phase separated regions of the phase diagram.

The inset to Fig. \ref{fig6} reports $\mu(n)$ for $a_1=0.7t$ revealing
that, for this parameter set and upon lowering the electron
density, $\kappa$ diverges at $n\approx 0.05$. However, the
global instability towards PS sets in already at larger
$n$ and is most conveniently extracted from the tangent construction
to the free energy, as exemplified in the main panel of Fig. \ref{fig6}.
Upon increasing the coupling $a_1$ the corresponding touching points   
move along the solid line labeled 'PS' in the lower right inset to 
Fig. \ref{fig7}.

For completeness, we point out that
upon further increasing $a_1$ and when the chemical potential
has entered the upper RSO-split band, the $\mu$ {\it vs.}
$n$ curve develops another minimum. At the transition the
inflection point with $\partial\mu/\partial n=0$ corresponds
to the divergent compressibility and ${\bf q}=0$ instability
which is reported by squares in Fig. \ref{fig1}.
In this high-density case, the
Maxwell construction requires two tangent lines to the
free energy which for $a_1=0.804 t$ are shown in Fig. \ref{fig7}.
Coming from large density the system tends to phase
separate between $n\approx 0.35$ (cf.
squares in Fig. \ref{fig1}) and the other touching
point of the tangent at $n\approx 0.18$. This density
marks the upper boundary of a tiny stable region which is
bound from below by the tangent which connects to $n=0$.
The PS line, which is shown in the upper right inset to 
Fig. \ref{fig7} together with the
instabilites from Fig. \ref{fig1}, 
does not comprise this situation but is limited to the
transition at small $a_1$.

\section{Discussion and Conclusions}\label{sec:discuss}
Our present investigations reveal the occurence of both PS 
and long-wavelength charge instabilities in a
spin-orbit coupled system where the RSO coupling depends
on the electron density (via the confining interface electric field). 
In particular, we have found
that within our model the global transition towards PS
is in close vicinity to a second order CDW instability so that
PS is accompanied by significant long-wavelength
charge  fluctuations. Our investigations generalize
the analysis of Refs. \cite{caprara12,bucheli132}, where
phases with negative compressibility have been evaluated for a
2D electron gas as realized in oxide heterostructures.

Typically these 2D electron gases are formed at the interfaces between 
two oxides consisting, respectively, of polar
[e.g., (LaO)$^+$ and (AlO$_2$)$^-$] and non-polar 
[e.g., TiO$_2$ and SrO] layers, with typical electron densities 
(per unit cell) $n\approx 0.01-0.05$. In this regard, a proper description
of the electronic states should at least take into account the multiband
structure due to the splitting of Ti$_{3d}$ t$_{2g}$ bands at the
interface. According to the analysis of Refs. 
\cite{caprara12,bucheli132} the system is in general stable when
the chemical potential falls in the 'light mass' lower isotropic band
of mainly $d_{xy}$ character whereas a PS instability
may occur when the gating potential shifts the Fermi energy into the higher anisotropic $d_{xz}$ and $d_{yz}$ bands.

Our present results refer to a model hamiltonian that can be considered as 
an effective model for these higher energy 'active' bands. In fact,
the RSO coupling  used in our analysis compares well with 
standard values (see. e.g. Ref. \cite{bucheli132}) 
when the hopping parameter is derived from the dispersion of 
these 'active'  bands \cite{note}.
The inset to Fig. \ref{fig6}
anticipates then the behavior of the chemical potential (dotted line) 
in case of an additional 'stable' lower energy band with 
$\partial\mu/\partial n >0$. This implies that the lower bound of the
phase separated regime as determined by a standard Maxwell construction
 is no longer at $n=0$ as in our one-band model:
 The tangent of the Maxwell construction would connect to a 
density in the 'stable' lower band.
On the other hand, it
is conceivable that the implementation of anisotropic 'active' bands
supports the occurence of CDW instabilities which then in Fig. \ref{fig1}
would still be the leading transition at small density.

It would therefore be interesting to explore possible realizations
of these instabilities in an 'unrestricted' multiband model. It might
well be that the combination of PS and long-wave length charge
instabilities leads to a kind of 'correlated disorder' which
is compatible with the existence of tails in the sheet
resistance curves of oxide interfaces below the superconducting transition
temperature \cite{bucheli13}. Work in this direction is in progress.
 
\acknowledgments 
We acknowledge insightful discussions with L. Benfatto, N. Bergeal, 
V. Brosco, C. Castellani, C. Di Castro,  J. Lesueur, and R. Raimondi. G.S.  
acknowledges support from the Deutsche Forschungsgemeinschaft. M.G. and S.C. acknowledge financial 
support from University Research Project of the University of Rome Sapienza, No. C26A125JMB.

\end{document}